\documentclass{PoS}
\usepackage{float} 
\usepackage{latexsym,amsmath,amstext}
\usepackage{physics}
\usepackage{amssymb,fontenc,times,mathptmx,graphicx}

\title{Static Potential At Non-zero Temperatures From Fine Lattices}

\author{
  \speaker{D. Hoying$^1$}, A. Bazavov$^{1}$, 
  D. Bala$^{2}$, G. Parkar$^{3}$, O. Kaczmarek$^2$, R. Larsen$^3$, Swagato Mukherjee$^4$, P. Petreczky$^4$, A. Rothkopf$^3$, J. H. Weber$^5$\\
\llap{$^1$}Department of Computational Mathematics, Science and Engineering, and 
Department of Physics and Astronomy, Michigan State University, East Lansing, MI 48824, USA\\
\llap{$^2$}Fakult\"at f\"ur Physik, Universit\"at Bielefeld, D-33615 Bielefeld, Germany\\
\llap{$^3$} Faculty of Science and Technology, University of Stavanger, NO-4036 Stavanger, Norway,\\
\llap{$^4$}Physics Department, Brookhaven National Laboratory, Upton, NY 11973, USA \\
\llap{$^5$} Institut f\"ur Physik \& IRIS Adlershof, Humboldt-Universit\"at zu Berlin, D-12489 Berlin, Germany
}

\abstract{We report on a preliminary study of static quark anti-quark potential at non-zero temperature in $2+1$ flavor QCD using 
$96^3\times N_{\tau}$ lattices with lattice spacing $a=0.028$fm, physical 
strange quark mass and light quark masses corresponding to pion mass of about $300$ MeV. We use $N_\tau=32,~24,~20$ and $16$
that correspond to temperature range $T=220-441$ MeV.
The in order to obtain the potential we calculate the Wilson line correlator in Coulomb gauge with additional HYP 
smearing to reduce the noise at large quark anti-quark separations. 
We apply $0$, $5$ and $10$ steps of HYP smearing to ensure that there is no physical effect from over-smearing. 
At the two highest temperatures we also consider a noise reduction technique that is based on an interpolation in the spatial separation between
the static quark and anti-quark.
}

\FullConference{%
 The 38th International Symposium on Lattice Field Theory, LATTICE2021
  26th-30th July, 2021
  Zoom/Gather@Massachusetts Institute of Technology
}

\ShortTitle{Potential At Non-zero Temperatures}

\begin{document}
\maketitle

\section{Introduction}

There has been a considerable interest in studying quarkonium properties at non-zero temperature since the
seminal paper of Matsui and Satz, that suggested that suppression of quarkonium production in heavy
ion collisions can signal creation of a deconfined medium \cite{Matsui:1986dk}. In-medium properties
of heavy quarkonium are encoded in the spectral functions, which in turn can be related to
Euclidean time meson correlation functions, see e.g. Ref. \cite{Bazavov:2009us} for a review.
Reconstruction of the quarkonium spectral functions from a discrete set of data points turned
out to be a very challenging task, see e.g. Refs. \cite{Wetzorke:2001dk,Karsch:2002wv,Datta:2003ww}.
At zero temperature quarkonium properties can be estimated quite well using potential models with
static quark antiquark potential
calculated in lattice QCD, see e.g. Ref. \cite{QuarkoniumWorkingGroup:2004kpm}.
It has been also proposed to calculate the quarkonium spectral function at non-zero temperature
using potential model with complex potential from lattice QCD \cite{Petreczky:2010tk,Burnier:2015tda}.
The complex potential at non-zero temperature is also important for real time modeling of 
quarkonium production in heavy ion collisions \cite{Yao:2021lus,Rothkopf:2019ipj}. 
The complex potential at non-zero temperature can be defined in terms of Wilson loops or correlators
of Wilson lines in Coulomb gauge, $W(r,\tau,T)$ through their spectral decomposition \cite{rothkopf2009proper,Rothkopf:2011db}
\begin{align*}
 W(r,\tau,T)&=\int_{-\infty}^\infty d\omega \rho_r(\omega, T)e^{-\omega\tau}.
\end{align*}
Here $r$ is the spatial separation between the static quark and antiquark and acts as label index
for the spectral function of static $Q\bar Q$ pair, $\rho_r(\omega, T)$.
If the spectral function has a peak at some value of $\omega$, the peak position gives the real part of
the potential
$ReV (r, T )$, while its  width gives the imaginary part of the potential, $ImV (r, T )$.
We still need to reconstruct the spectral function in order to determine the potential, but
the structure of this spectral function is much simpler than the structure of quarkonium spectral function, and thus should be easier to reconstruct.
One needs a large number of data points in the time direction to accomplish this task.
Current state of the art calculation in 2+1 flavor QCD mostly use 
lattices with temporal extent $N_\tau = 12$ and some $N_{\tau}=16$ lattices that are available only at high temperatures \cite{nt12pap}.
In this contribution we report preliminary calculations of the Wilson line correlators
calculated on very fine lattices with lattices spacing $a^{-1} = 7.04$ GeV and temporal
extent up to $N_{\tau}=32$.

\section{Lattice setup}
We perform calculations at non-zero temperature in 2+1 flavor QCD on $96^3\times N_\tau$ lattices
with lattice spacing $a^{-1} = 7.04$ GeV. The strange quark mass, $m_s$ is fixed to its physical value,
while for the light quark mass we use $m_l=m_s/5$, which corresponds to the pion mass of about $300$ MeV
in the continuum limit.
We consider 
$N_\tau = 32,~24,~20$ and $16$, which correspond to $T = 220,~294,~353$ and $441$ MeV, respectively.
We also generated 
additional $T = 0$ ($64^4$ ) lattices to serve as a reference.
We use temporal Wilson line correlators in Coulomb gauge instead of Wilson
loops to obtain a better signal for the potential. However, this approach turns out to be insufficient
for our very fine lattices. Therefore, the temporal Wilson lines are constructed from HYP smeared \cite{Hasenfratz:2001hp}
temporal gauge links for $N_{\tau}=32$ and $24$, as well as for the zero temperature case.
We use 5 and 10 steps of HYP smearing. For the two highest temperature we consider an alternative
noise reduction approach, which will be discussed below.

\section{Effective Masses}

\begin{figure}
\includegraphics[width=8cm]{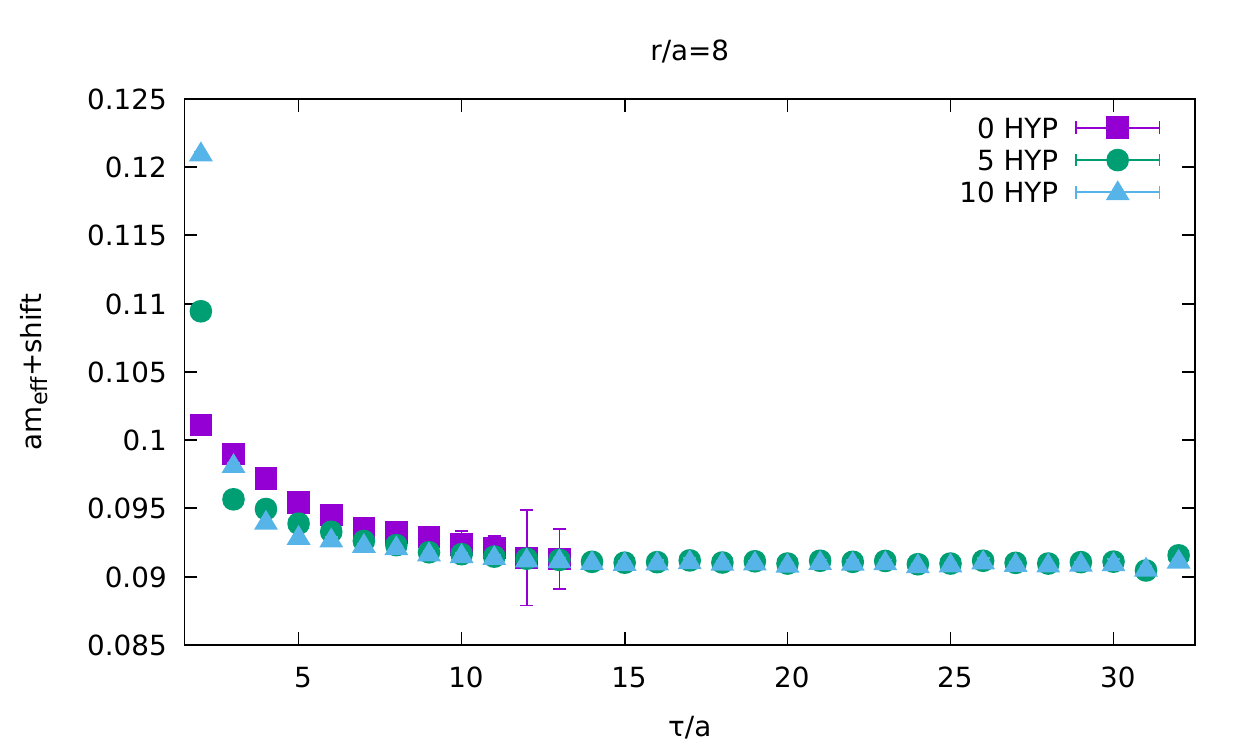}
\caption{The effective masses for Wilson line correlator for $r/a=8$ for 0, 5 and 10 steps of HYP smearing.}
\label{fig:meff0}
\end{figure}

Before reconstructing the spectral function $\rho_r(\omega,T)$ we need to understand
the $\tau$-dependence of the Wilson line correlators at different temperatures. This can be done
in terms of the effective mass
\begin{align*}
  m_{\text{eff}}(r,\tau, T)&=\partial_\tau \ln W(r,\tau, T)
  \\ &\simeq \frac{1}{a}\ln \frac{W(r,\tau, T)}{W(r,\tau+a, T)}.
\end{align*}
At zero temperature the effective mass approaches a plateau that corresponds to the ground state energy, i.e.
the static potential, $V(r)$. Our results for the effective masses at zero temperature are shown in Fig. \ref{fig:meff0}.
Since the potential has a divergent part proportional to the inverse lattice spacing and the coefficient
of this divergence depends on the number of smearing steps, the effective masses have been shifted by a
constant to match them in the plateau region. As one can see from the figure for $\tau/a>3$ more smearing
steps leads to faster approach to the plateau. However, for the two smaller $\tau$ values this is not the case,
and for the smallest $\tau$, the effective mass is the largest for 10 steps of HYP smearing.
Without HYP smearing we loose the signal for $\tau/a \gtrsim 12$ and therefore, we do not show the corresponding results.
At larger spatial separations the situation is even worse, and no hint of a plateau can be obtained from
unsmeared results. The HYP smearing distorts the $r$-dependence of the static potential. But these distortions
are limited to small distances. For distances $r/a>5$, which are relevant for our study, the distortions due to
HYP smearing are negligible compared to other sources of errors.

\begin{figure}
\includegraphics[width=7cm]{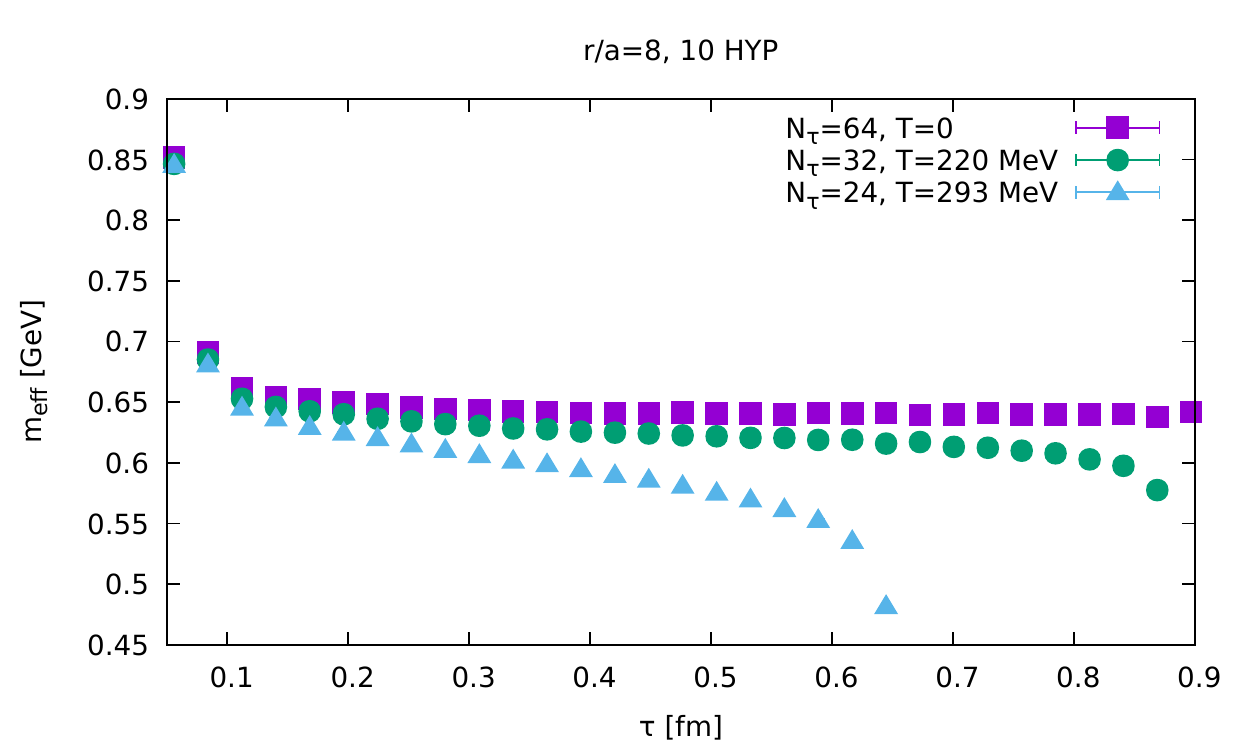}
\includegraphics[width=7cm]{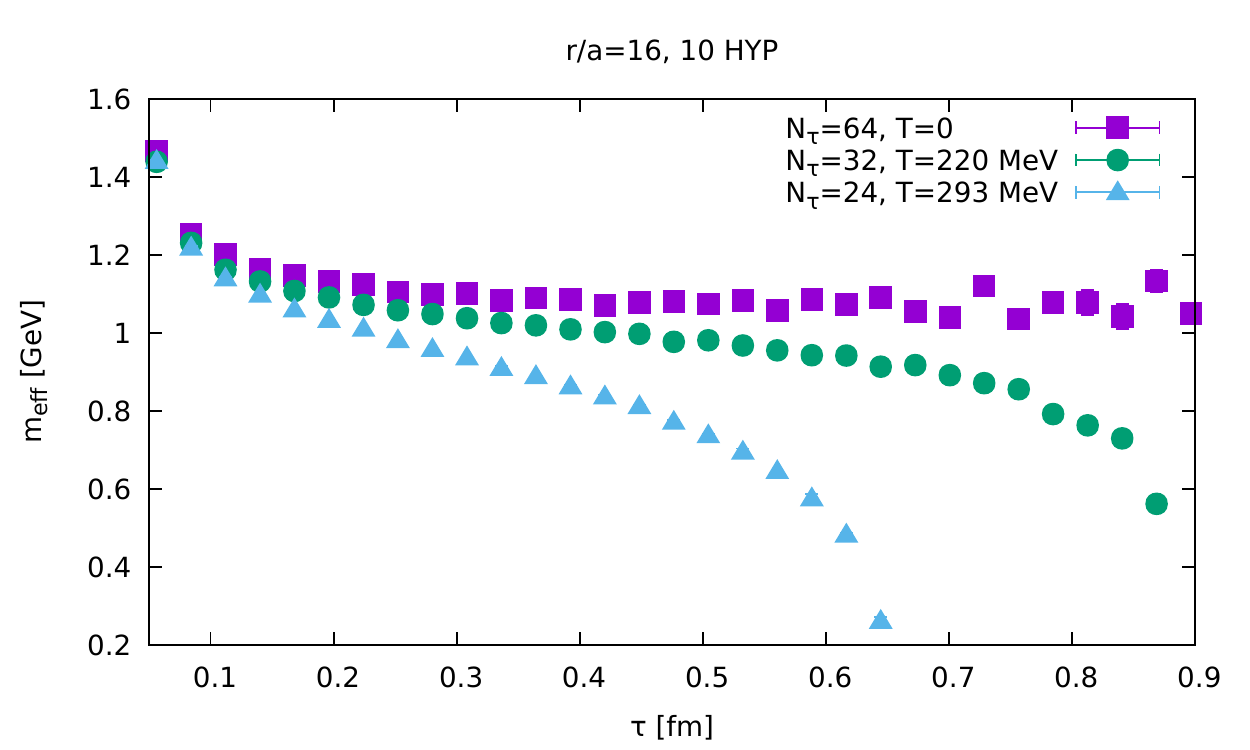}
\caption{The effective masses at different temperature as function of $\tau$ for $r/a=8$ (left) and
$r/a=16$ (right). The results for 10 steps of HYP smearing are shown.}
\label{fig:meffTdep}
\end{figure}

Next, we examine the temperature dependence of the effective masses. In Fig. \ref{fig:meffTdep} we show 
the effective mass for $T=0$, $T=220$ MeV and $T=293$ MeV at two different distances, $r/a=8$ and $r/a=16$
and 10 steps of HYP smearing. The results for 5 steps of HYP smearing are similar.
We see that for small $\tau$ the temperature effects are quite small and grow with increasing $\tau$.
At non-zero temperature the effective masses do not show a plateau at large $\tau$ but an approximately linear
decrease followed by a very rapid strongly non-linear drop as $\tau$ approaches $1/T$. These features follow
from the general properties of the spectral function $\rho_r(\omega,T)$ \cite{nt12pap}. 
The approximately linear decrease in the effective masses is due to the fact that the ground
state acquires a thermal width. 
The sharp drop at large $\tau$ is caused by the low $\omega$ tail of the broadened peak \cite{nt12pap}.
The thermal effects are obviously larger at larger separations.

\begin{figure}
\includegraphics[width=7cm]{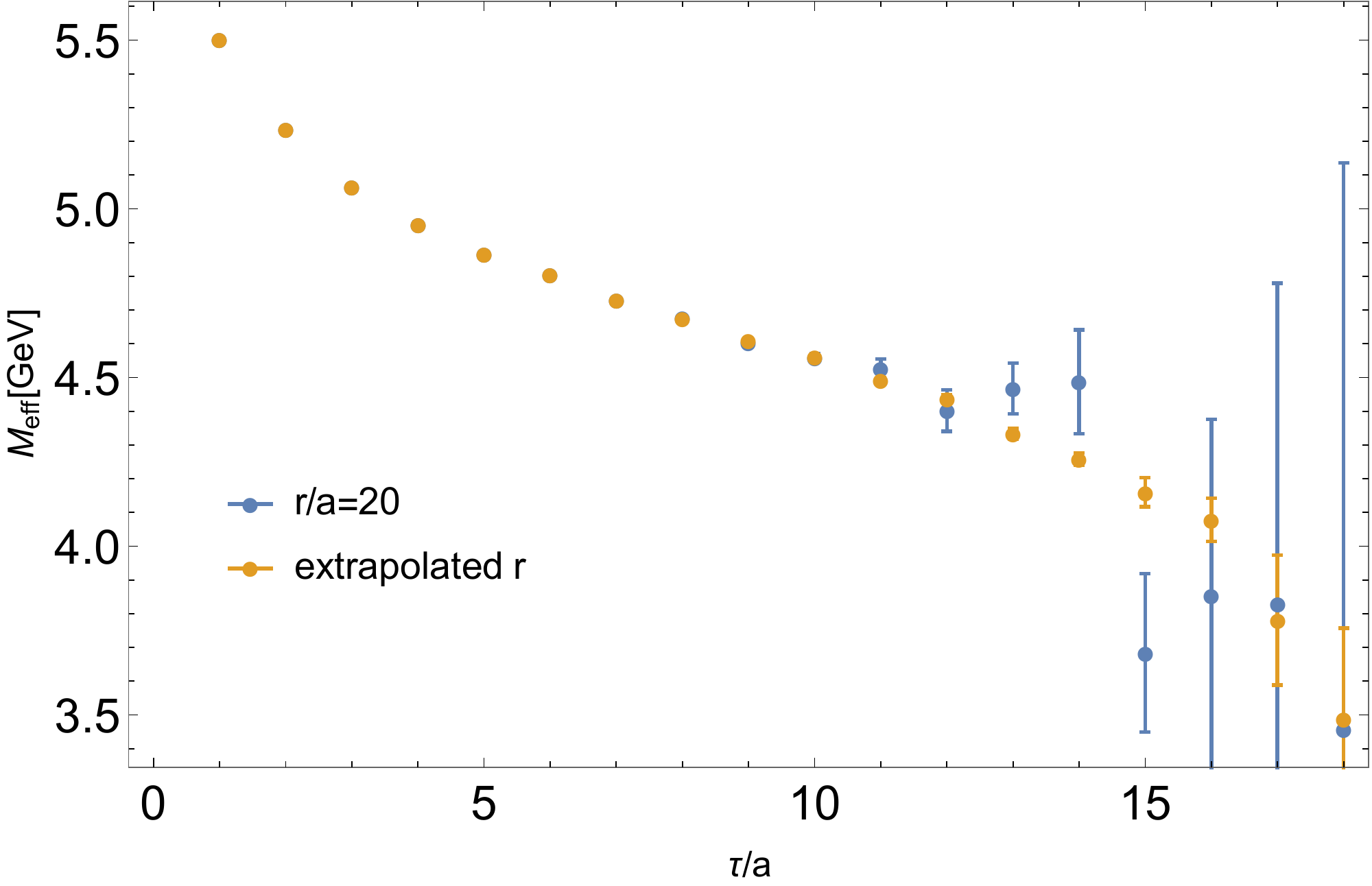}
\includegraphics[width=7cm]{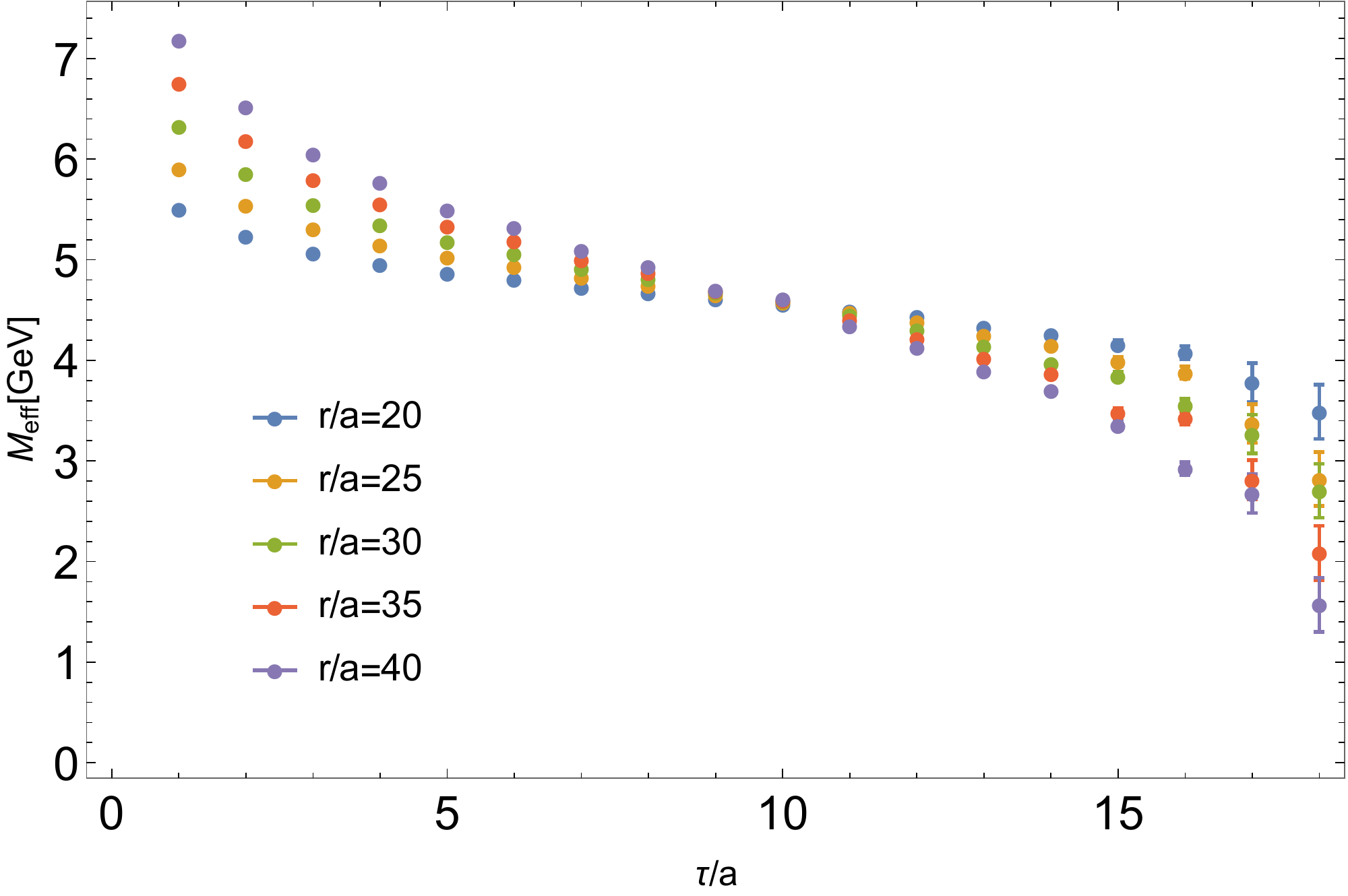}
\caption{The effective masses for $T=353$ MeV ($N_\tau=20$) for different separation.
In the left panel we show the effective mass obtained using interpolation in $r$ and compared
to the standard result on the effective mass. In the right panel we show results obtained
form the interpolation procedure for $r/a=20,~25,~30,~35$ and $40$.}
\label{fig:meffnt20}
\end{figure}

For $N_\tau=20$ we tried another procedure for noise reduction. It is based on the idea that 
for any fixed $\tau$ the Wilson line correlator is a smoothly decaying function of $r$.
The noise problem shows up at large values of $r$, where the correlation
functions is available for many different spatial separations around a particular $r$ value, that differ by fraction of
the lattice spacing.
Since the correlation function should be smooth in $r$ (lattice artifacts at large distances are negligible)
we can reduce the fluctuations
in the data set by performing 
smooth interpolation in $r$ in a narrow region of $r$.
We used second order polynomial interpolations in intervals of $r/a$ that are smaller than $0.9$ for $r/a\ge 20$.
The statistical fluctuations are largely reduced as the result of these interpolations.
Using the corresponding interpolations for each $\tau$
we can calculate the effective mass using the smoothened data for prescribed $r$ values.
In Fig. \ref{fig:meffnt20} we show the result of such an analysis. We indeed see that the interpolation
in $r$ really helps to reduce the statistical noise in the correlator. The right panel of this 
figure shows that the procedure is very effective for large $r$ values, $r/a\ge 20$.

\subsection{Subtracted correlators and comparison with the previous results}
In the previous section we have seen that at very small $\tau$ 
the temperature dependence of 
the Wilson line
correlators is very small. This is expected to be due to the fact that the at small Euclidean
time the correlation function mostly receives contributions from the high $\omega$ part of the spectral
function, and this part of the spectral function is largely temperature
independent, see e.g. Ref. \cite{Bazavov:2009us}.
\begin{figure}
\includegraphics[width=7cm]{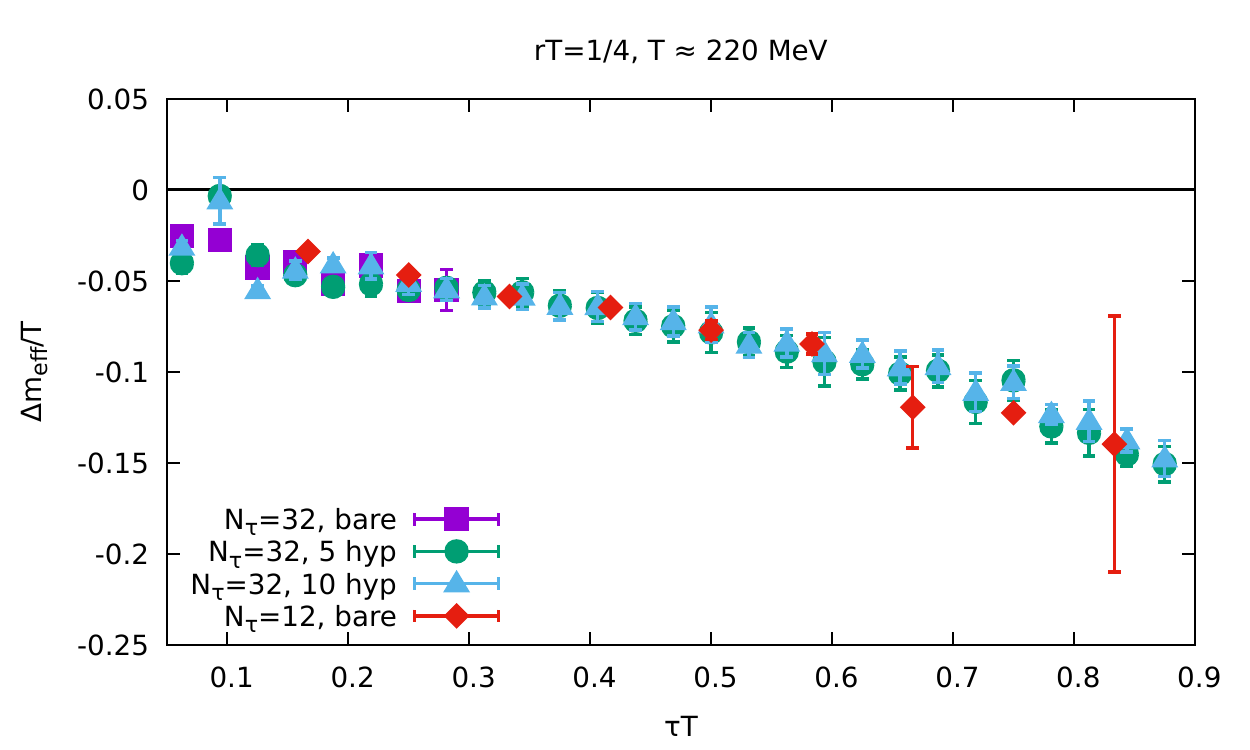}
\includegraphics[width=7cm]{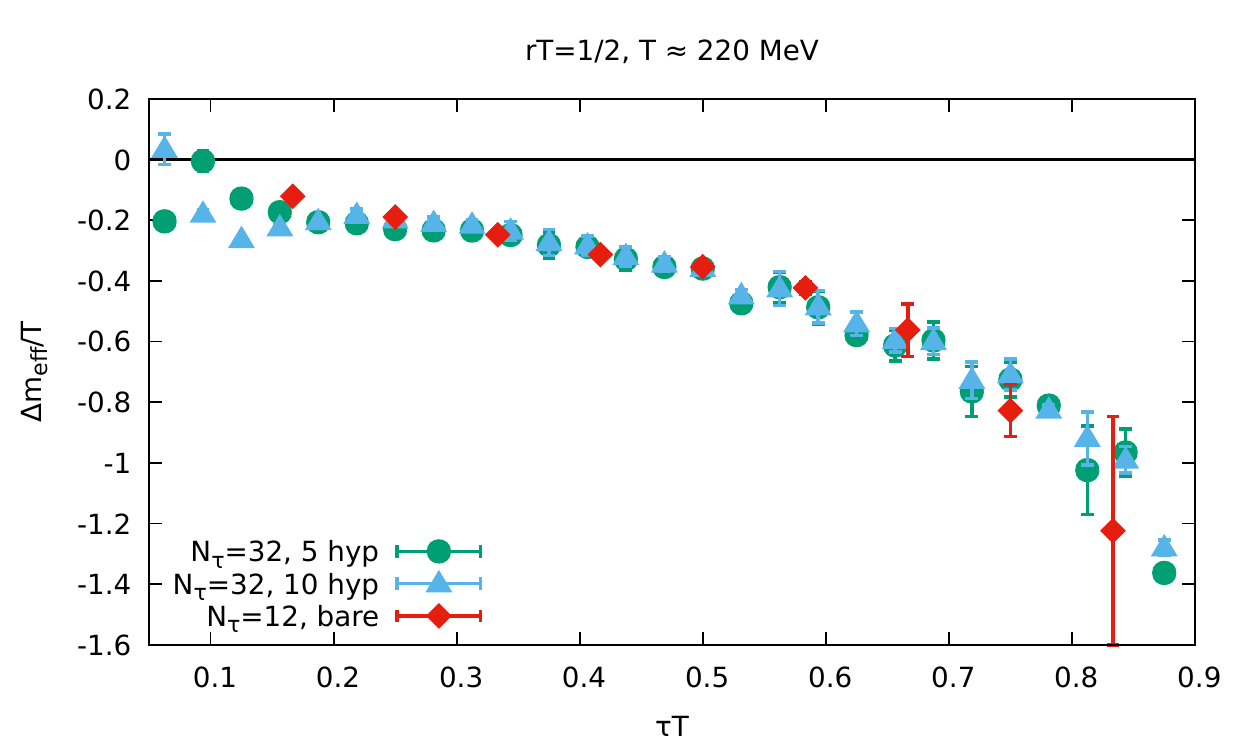}

\includegraphics[width=7cm]{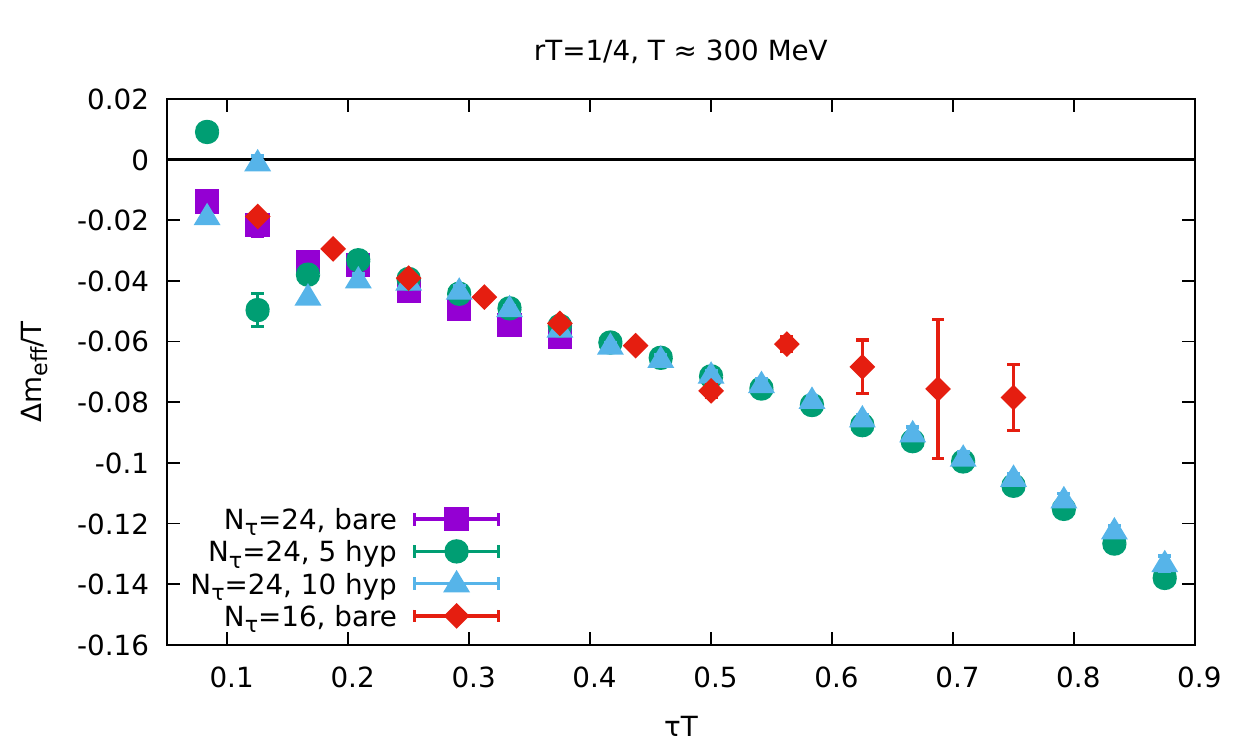}
\includegraphics[width=7cm]{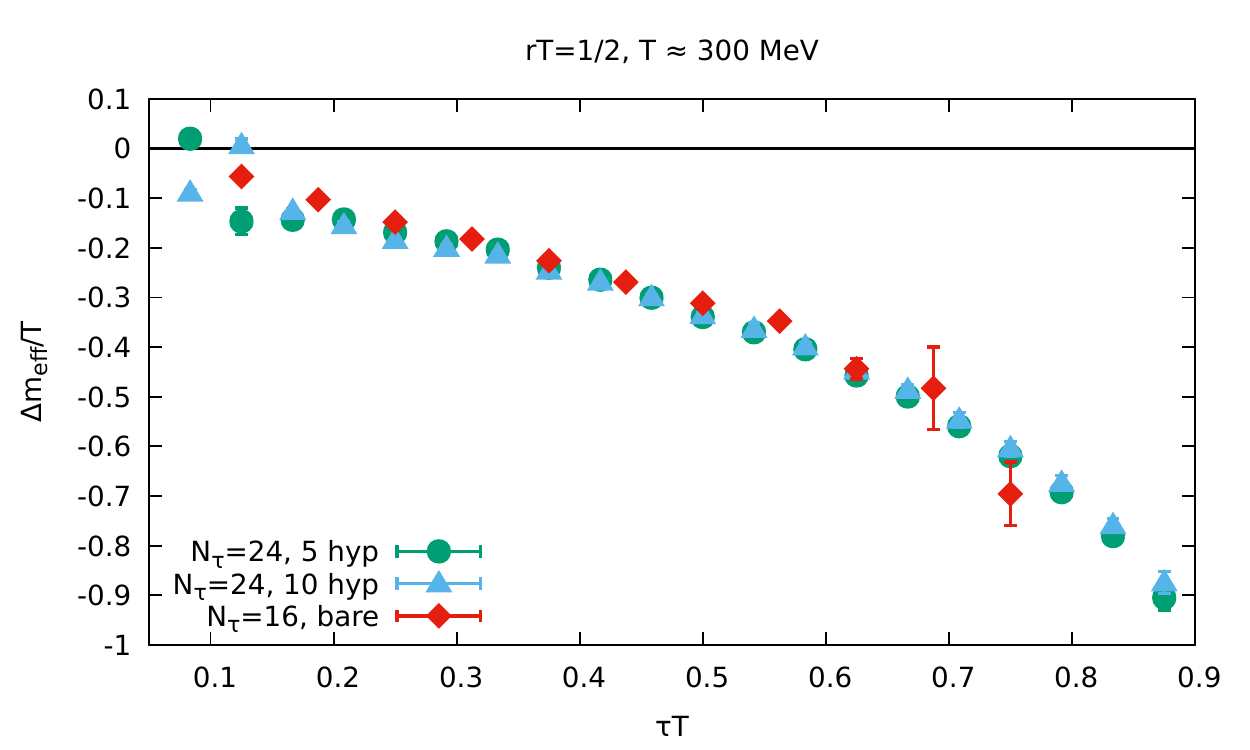}
\caption{The effective masses corresponding to the subtracted correlators at two distances:
$rT=1/4$ (left) and $rT=1/2$ (right). The top panel shows the results for $T=220$ MeV, while
the bottom panels show the results for $T=300$ MeV. We compare our results with the previous
$N_{\tau}=12$ and $N_{\tau}=16$ results without smearing, i.e. bare results.}
\label{fig:meff_subtr}
\end{figure}
Therefore, for the spectral function we can write \cite{Larsen:2019bwy,Larsen:2019zqv}
\begin{equation}
 \rho_r(\omega, T)=\rho_r^{\text{peak}}(\omega, T)+ \rho_r^{\text{high}}(\omega),
\end{equation}
with $\rho^{peak}(\omega,T)$ corresponding to the ground state peak in the spectral function that is broadened at
non-zero temperature and may have a large low $\omega$ tail \cite{nt12pap}.
This equation implies that
\begin{equation}
W(r,\tau,T)=W^{\text{peak}}(r,\tau,T)+W^{\text{high}}(r,\tau).
\end{equation}
The zero temperature spectral functions has a delta peak corresponding to the ground state,
$\rho_r(\omega, T=0)=A \delta(\omega-V^{T=0}(r))+\rho_r^{\text{high}}(\omega)$. Therefore,
by fitting the ground state contribution at zero temperature and then subtracting it from
$W(r,\tau,T=0)$ we can estimate $W^{\text{high}}(r,\tau)$. If we subtract $W^{\text{high}}(r,\tau)$
from the finite temperature Wilson line correlator we obtain the subtracted correlator that is mainly
sensitive to the temperature dependent peak part of the spectral function, $\rho_r^{\text{peak}}(\omega, T)$.
Therefore, it is important to analyze the effective masses for the subtracted correlator. The
corresponding results are shown in Fig. \ref{fig:meff_subtr}. We also compare our results with
the ones obtained on $N_{\tau}=12$ and $N_{\tau}=16$ lattice without HYP smearing \cite{nt12pap}.
The effective masses show a linear decrease in $\tau$ when the Euclidean time separation is far away from $1/T$.
For $\tau$ close to $1/T$ we see a faster non-linear dependence. These features of the subtracted
effective masses are also present in the unsmeared $N_{\tau}=12$ and $N_{\tau}=16$ data and we find
a very good agreement with the corresponding results, see Fig. \ref{fig:meff_subtr}.
For large time the new results have significantly smaller errors as the result of HYP smearing.
The significant smearing dependence of the effective masses at small $\tau$ is largely reduced due to the subtraction, i.e.
HYP smearing was mostly affecting the high energy part of the spectral function. The subtracted 
effective masses at small $\tau$ show some non-monotonic behavior at very small $\tau$. This 
is due to the distortions due to HYP smearing, which affect slightly differently the zero
and finite temperature correlator. The unsmeared data, on the other hand, show the expected behavior.

\section{Conclusion}
To obtain the the complex static $Q \bar Q$ potential at non-zero temperature one needs to
calculate Wilson loops or Wilson line correlators on fine lattices with large temporal extent.
This, however, is challenging because of the poor signal to noise ratio. Therefore, in this contribution
we explored two possible avenues for noise reduction. The first one is to use HYP smearing on the 
temporal links. We observed that using 5 and 10 steps of HYP smearing on lattices with $a^{-1}=7.04$ GeV 
at $T=220$ MeV and $T=330$ MeV provides a good signal for all relevant $Q\bar Q$ separations.
The distortions due to HYP smearing are limited to very small $\tau$, namely $\tau/a<5$. Therefore,
HYP smearing is a viable strategy for noise reduction. We also explored interpolation in $r$ to 
reduces the noise. This approach worked well at the two highest temperature, $T=353$ and $441$ MeV.
The subtracted effective masses agree well with the previous calculations performed on $N_{\tau}=12$
and $N_{\tau}=16$ lattices but have much smaller errors for large $\tau$ values. Using the presented 
data on the effective masses of the Wilson line correlators we can obtain the complex static $Q\bar Q$ potential
if a suitable Ansatz for the spectral function is introduced. The corresponding analysis is currently underway.

\section*{Acknowledgement}
This material is based upon work supported by the U.S. Department of Energy, Office of Science, 
Office of Nuclear Physics: (i) Through the Contract No. DE-SC0012704; 
(ii) Through the Scientific Discovery through Advance Computing (SciDAC) award 
Computing the Properties of Matter with Leadership Computing Resources. 
R.L., G.P. and A.R. acknowledge funding by the Research Council of Norway under the FRIPRO Young Research Talent grant 286883. Part of the data analysis has been carried out on computing resources provided by UNINETT Sigma2 - the National Infrastructure for High Performance Computing and Data Storage in Norway under project NN9578K-QCDrtX "Real-time dynamics of nuclear matter under extreme conditions". 
J.H.W.’s research was funded by the Deutsche Forschungsgemeinschaft (DFG, German Research Foundation) - Projektnummer 417533893/GRK2575 ``Rethinking Quantum Field Theory''. 
D.B. and O.K. acknowledge support by the Deutsche Forschungsgemeinschaft (DFG, German Research Foundation) through the CRC-TR 211 'Strong-interaction matter under extreme conditions'– project number 315477589 – TRR 211.
This research used resources of the National Energy Research Scientific Computing Center (NERSC), a U.S. Department of Energy Office of Science User Facility located at Lawrence Berkeley National Laboratory, operated under Contract No. DE-AC02-05CH11231.
Some of the  numerical calculations have been performed on JUWELS Booster  J\"ulich Supercomputing Center using allocation from PRACE.

\bibliographystyle{JHEP}
\bibliography{lat21.bib}

\providecommand{\href}[2]{#2}\begingroup\raggedright\begin{thebibliography}{10}

\bibitem{Matsui:1986dk}
T.~Matsui and H.~Satz, \emph{{$J/\psi$ Suppression by Quark-Gluon Plasma
  Formation}}, \href{https://doi.org/10.1016/0370-2693(86)91404-8}{\emph{Phys.
  Lett. B} {\bfseries 178} (1986) 416}.

\bibitem{Bazavov:2009us}
A.~Bazavov, P.~Petreczky and A.~Velytsky, \emph{{Quarkonium at Finite
  Temperature}}, pp.~61--110.
\newblock 2010.
\newblock \href{https://arxiv.org/abs/0904.1748}{{\ttfamily 0904.1748}}.

\bibitem{Wetzorke:2001dk}
I.~Wetzorke, F.~Karsch, E.~Laermann, P.~Petreczky and S.~Stickan, \emph{{Meson
  spectral functions at finite temperature}},
  \href{https://doi.org/10.1016/S0920-5632(01)01763-7}{\emph{Nucl. Phys. B
  Proc. Suppl.} {\bfseries 106} (2002) 510}
  [\href{https://arxiv.org/abs/hep-lat/0110132}{{\ttfamily hep-lat/0110132}}].

\bibitem{Karsch:2002wv}
F.~Karsch, S.~Datta, E.~Laermann, P.~Petreczky, S.~Stickan and I.~Wetzorke,
  \emph{{Hadron correlators, spectral functions and thermal dilepton rates from
  lattice QCD}},
  \href{https://doi.org/10.1016/S0375-9474(02)01470-7}{\emph{Nucl. Phys. A}
  {\bfseries 715} (2003) 701}
  [\href{https://arxiv.org/abs/hep-ph/0209028}{{\ttfamily hep-ph/0209028}}].

\bibitem{Datta:2003ww}
S.~Datta, F.~Karsch, P.~Petreczky and I.~Wetzorke, \emph{{Behavior of
  charmonium systems after deconfinement}},
  \href{https://doi.org/10.1103/PhysRevD.69.094507}{\emph{Phys. Rev. D}
  {\bfseries 69} (2004) 094507}
  [\href{https://arxiv.org/abs/hep-lat/0312037}{{\ttfamily hep-lat/0312037}}].

\bibitem{QuarkoniumWorkingGroup:2004kpm}
{\scshape Quarkonium Working Group} collaboration, N.~Brambilla et~al.,
  \emph{{Heavy quarkonium physics}},
  \href{https://arxiv.org/abs/hep-ph/0412158}{{\ttfamily hep-ph/0412158}}.

\bibitem{Petreczky:2010tk}
P.~Petreczky, C.~Miao and A.~Mocsy, \emph{{Quarkonium spectral functions with
  complex potential}},
  \href{https://doi.org/10.1016/j.nuclphysa.2011.02.028}{\emph{Nucl. Phys. A}
  {\bfseries 855} (2011) 125}
  [\href{https://arxiv.org/abs/1012.4433}{{\ttfamily 1012.4433}}].

\bibitem{Burnier:2015tda}
Y.~Burnier, O.~Kaczmarek and A.~Rothkopf, \emph{{Quarkonium at finite
  temperature: Towards realistic phenomenology from first principles}},
  \href{https://doi.org/10.1007/JHEP12(2015)101}{\emph{JHEP} {\bfseries 12}
  (2015) 101} [\href{https://arxiv.org/abs/1509.07366}{{\ttfamily
  1509.07366}}].

\bibitem{Yao:2021lus}
X.~Yao, \emph{{Open quantum systems for quarkonia}},
  \href{https://doi.org/10.1142/S0217751X21300106}{\emph{Int. J. Mod. Phys. A}
  {\bfseries 36} (2021) 2130010}
  [\href{https://arxiv.org/abs/2102.01736}{{\ttfamily 2102.01736}}].

\bibitem{Rothkopf:2019ipj}
A.~Rothkopf, \emph{{Heavy Quarkonium in Extreme Conditions}},
  \href{https://doi.org/10.1016/j.physrep.2020.02.006}{\emph{Phys. Rept.}
  {\bfseries 858} (2020) 1} [\href{https://arxiv.org/abs/1912.02253}{{\ttfamily
  1912.02253}}].

\bibitem{rothkopf2009proper}
A.~Rothkopf, T.~Hatsuda and S.~Sasaki, \emph{Proper heavy-quark potential from
  a spectral decomposition of the thermal wilson loop},  2009.

\bibitem{Rothkopf:2011db}
A.~Rothkopf, T.~Hatsuda and S.~Sasaki, \emph{{Complex Heavy-Quark Potential at
  Finite Temperature from Lattice QCD}},
  \href{https://doi.org/10.1103/PhysRevLett.108.162001}{\emph{Phys. Rev. Lett.}
  {\bfseries 108} (2012) 162001}
  [\href{https://arxiv.org/abs/1108.1579}{{\ttfamily 1108.1579}}].

\bibitem{Hasenfratz:2001hp}
A.~Hasenfratz and F.~Knechtli, \emph{{Flavor symmetry and the static potential
  with hypercubic blocking}},
  \href{https://doi.org/10.1103/PhysRevD.64.034504}{\emph{Phys. Rev. D}
  {\bfseries 64} (2001) 034504}
  [\href{https://arxiv.org/abs/hep-lat/0103029}{{\ttfamily hep-lat/0103029}}].

\bibitem{nt12pap}
D.~Bala et~al., \emph{Static quark anti-quark interactions at non-zero
  temperature from lattice qcd, preprint hu-ep-21/32-rtg}, .

\bibitem{Larsen:2019bwy}
R.~Larsen, S.~Meinel, S.~Mukherjee and P.~Petreczky, \emph{{Thermal broadening
  of bottomonia: Lattice nonrelativistic QCD with extended operators}},
  \href{https://doi.org/10.1103/PhysRevD.100.074506}{\emph{Phys. Rev. D}
  {\bfseries 100} (2019) 074506}
  [\href{https://arxiv.org/abs/1908.08437}{{\ttfamily 1908.08437}}].

\bibitem{Larsen:2019zqv}
R.~Larsen, S.~Meinel, S.~Mukherjee and P.~Petreczky, \emph{{Excited bottomonia
  in quark-gluon plasma from lattice QCD}},
  \href{https://doi.org/10.1016/j.physletb.2019.135119}{\emph{Phys. Lett. B}
  {\bfseries 800} (2020) 135119}
  [\href{https://arxiv.org/abs/1910.07374}{{\ttfamily 1910.07374}}].

\end{thebibliography}\endgroup




\end{document}